# Meta-separation: complete separation of organic-water mixtures by structural property of metamaterial

**Kyoung Hwan Choi[1], Da Young Hwang[1], Jeong Eon Park[1] and Dong Hack Suh[1]†**

*1 Advanced Materials & Chemical Engineering Building 311, 222 Wangsimni-ro, Seongdong-Gu, Seoul, Korea, E-mail: dhsuh@hanyang.ac.kr*

**Abstract**

The separation of liquid mixture has been studied for a long time. Separation proceeds based on the difference in physical properties including pore size and electrostatic interaction. Therefore, there are many difficulties in separation of materials having similar size or polarities in physical properties such as ethanol-water and 1,4-dioxane-water mixtures. While we still lack a universal generalization of these ideas to the separation, pervaporation based on a difference in transport rates by permeability through a membrane by the permeate was early suggested. Yet there is an existing technical gap to remove trace amounts of organics dissolved in water. Here, we report a novel separation strategy employing a metamaterial, called meta-separation using the exotic structural property of metamateirals rather than electrostatic characteristics. The structural properties of metamaterials provide various functions of super-hydrophobicity based on roughness of surface, the strong capillary effect based on nanopore, and huge void for great absorption of organics. It exhibited a water contact angle of 151.3° and high adhesive property from nanopore. On the other hands, ethanol was immediately absorbed up to 93 wt%. This differences made it possible to quickly and easily eliminate organics dissolved in water. Furthermore, their applications are expected to achieve functions in environmental remediation, biofuel separation process, etc., without large scale facilities.

**Introduction**

Controlling the wettability of solid surface has recently attracted great interest from both fundamental and practical perspective by tailoring surface topography.[1-4] Inspired by the microstructure of lotus leaf which forms superhydrophobic surface in nature,[5] various superhydrophobic surfaces with different adhesive property can be controlled from periodicity, pore size and shape of nanopore.[6,7] Adhesive force of nanopore is originated from capillary force of it.[8]

On the other hands, nanopore has widely been used for various applications. Above all, gas adsorption due to the large surface area formed by nanopores has been importantly developed.[9-11] In addition, nanopore is actively used in various kinds of water treatment such as water/oil separation, desalination and water remediation.[12-14] In this case, most of them usually use only the pore size and their electrostatic characteristics on the surface. Therefore, it is difficult to find a way to use nanopore to separate mixtures of similar polarity and size.

To overcome these limitations, pervaporation employs a membrane for liquid separation, where a polymeric or zeolite membrane usually serves as the separating barrier.[15-19] This process provides incomparable advantages for the separation of azeotropic mixtures due to its mild operating conditions, and lack of environmental discharge for additional species in the feed stream.[20-23] Thus far, pervaporation has found viable application in the following three areas; dehydration of organic solvents, removal of dilute organic compounds from aqueous streams, and separation of organic-organic mixtures. Although the dehydration of organic solvents is a well-developed process, removing organic compounds from water still has certain limitations.[24]

Here in, we show exotic separation based on metamaterial. The superhydrophobic surface, originated from the surface roughness, makes barrier for water. In contrary, the capillary force originated from the huge void of bulk and nanopore of surface absorbs organics. A separation method based on both the surface and bulk properties shows a new separation method that is completely different from the size exclusion and electrostatic force based separation strategy. Moreover, unlike conventional membranes for separation, which have been provided a passive separation barrier against external driving force, the new method provides an active driving force itself by capillary force.

**Results and Discussion**

HYLION-12 with the building block for the Dirac metamaterial and metasurface was designed with two pairs of parallel strands of alkyl chains for the formation of the xy-plane and a pyrene moiety for stacking along the z-axis. This procedure is to fabricate the three-dimensional porous structure assembled by non-polar, non-covalent physical bonds.[25-27] Herein, it has two different structures which are orthorhombic and monoclinic phases designated as $\Phi_o$ and $\Phi_m$, respectively. Although $\Phi_m$ is usual molecular crystals, $\Phi_o$ shows not only exotic optical properties but also whole connected three dimensional porous structure. A lot of porous materials increase their surface area for gas separation through adsorption. $\Phi_o$ contrastively has minimized surface area and form the largest void inside. It hardly makes adsorption for gas on surface but can absorb liquid into wide inner pores. In order to utilize this characteristic for absorption, HYLION-14 with higher thermal stability was synthesized **(Fig. S1–S3)** and crystallized with various solvents. **(Table. S1)**. It was verified that HYLION-14 exhibits two different crystal morphologies, the needle- and the plate-types, when crystallized with ethyl acetate and benzene, respectively.

Powder X-ray diffraction (PXRD) was used to identify the crystal structure of HYLION-14 crystallized from ethyl acetate. The results reveal a highly ordered crystalline structure. At glance, the XRD pattern of needle type crystal of HYLION-14 is very similar to $\Phi_o$. However, it looks like that peaks at 2.35°, 4.68° and 6.92° which are originated from (100), (200) and (220) in $\Phi_o$ were shifted to a lower theta/2theta position. **(Fig. 1a)** This indicates that the unit cell size in both the x and y directions is longer by increasing the number of carbon atoms. The structure of needle type crystals was demonstrated for further analysis. The Miller indices were extracted from the PXRD peaks by employing n-TREOR09[28]. The crystal structure of HYLION-14 was solved via a direct method using EXPO2013[29,30], and the solutions of the structures were then refined by the Rietveld method[31]. Finally, the structure was modeled by using Discovery Studio[32]. The orthorhombic structure was designated as $\Phi_{o,14}$, belonging to the P2221 space group. The cell parameters of $\Phi_{o,14}$ were refined to a = 42.913 Å,

b = 45.996 Å, and c = 9.270 Å (residuals: $R_e$ = 8.532, $R_{wp}$ = 11.747, and $\chi^2$ = 1.37). **(Fig. 1b)** The XRD pattern of the plate-like crystal was solved in the same way as for $\Phi_{o,14}$. It adopts a primitive monoclinic crystal structure designated as $\Phi_{m,14}$, classified in the P2 space group with cell parameters of a = 9.024 Å, b = 43.341 Å, c = 8.303 Å, and β = 105.30° (residuals: $R_e$ = 8.257, $R_{wp}$ = 11.242, and $\chi^2$ = 1.36) and the unit cell of $\Phi_{o,14}$ is 151% larger than that of $\Phi_o$.

Differential scanning calorimetry (DSC) with temperature ramping from 45 °C to 95 °C at a heating rate of 1 °C min$^{-1}$ revealed two peaks at 74.01 and 81.11 °C, corresponding to phase transitions for $\Phi_{o,14}$. and an exothermic peak for crystallization at 70.60 °C upon cooling to 45 °C at a rate of 1 °C min$^{-1}$. **(Fig. 2a and b)** The first heating cycle of $\Phi_{o,14}$ had a similar profile with $\Phi_o$, but the transition temperatures increased by 13.65 °C. In addition, the first and the second phase transition energies have been increased from 10.05 to 11.70 kcal mol$^{-1}$ and 26.59 to 29.11 kcal mol$^{-1}$, respectively. It can be understood through the effect of van der Waals interaction energy difference with increasing the number of carbons. The second heating cycle curve showed only the second phase transition, which implies that the phase formed after annealing from the isotropic liquid is different from $\Phi_{o,14}$. The second heating cycle shows only one phase transition at 81.11°C. and it means that $\Phi_{o,14}$ is a metastable phase. The phase transition was investigated in depth by using high-temperature XRD (HT-XRD). The theta/2theta peaks at 9.87, 10.73, 11.48, and 19.48° in the profile of $\Phi_{o,14}$ were indexed to the (400), (330), (240), and (002) faces, respectively. **(Fig. 1c)** After heating to 74.01°C, the peak at 10.25° appeared, providing meaningful and important information of the phase transition. This can be interpreted as a change in the d-space of the (010) plane from 3.58 to 0.8 nm. During the phase transition, HYLION-14 in the middle layer of the unit cell of $\Phi_{o,14}$ rotates 90° from the x-axis to the y-axis. As originally intended, the structure with larger inner pores and higher termal stability is obtained from HYLION-14.

The N$_2$ gas adsorption of $\Phi_{o,14}$ depicts the porosities of it. $\Phi_{o,14}$ produced the nitrogen adsorption and desorption isotherms of a typical type IV with H3 hysteresis loops, according to the International Union of Pure and Applied Chemistry (IUPAC) classification. **(Fig. 3a)** This behavior indicates the predominance of mesopores.[33,34] Brunauer-Emmett-Teller (BET) surface area measurements and the *t*- plot analysis were carried out to determine the specific surface area of the as-prepared material. The BET surface area plot of $\Phi_{o,14}$ corresponded to the data from the BET equation.[35] **(Fig. 3b)** The specific surface areas of $\Phi_{o,14}$ were found to be 45.25 m$^2$ g$^{-1}$. These small surface areas indicate that alkyl chains of $\Phi_{o,14}$ is only suitable sites for adsorption.[36] It represents a plot of the volume of adsorbed (*Q*) nitrogen at different relative pressures against the thickness of the adsorbed layer, *t*, for $\Phi_{o,14}$. **(Fig. 3c)** The average pore diameter from the Barrett-Joyner-Halenda (BJH) desorption analysis was 11.3 nm, with a very wide pore size distribution, and the corresponding single-point total pore volume at *P/P$_0$* = 0.975 was 0.05 cm$^3$ g$^{-1}$. The experimental points in the *t*-plot method agreed with the data from the Harkins-Jura isotherm equation. It is clear from the plot that a set of experimental data points falls in a straight line for *t* = 0.35–0.45 nm. Thus, $\Phi_{o,14}$ was confirmed to be porous in nature, as the *t*-plot curves did not pass through the origin. The intercept of the fitted linear line was positive, which confirms the presence of mesopores.

The surface area of 45.25 m$^2$/g means that it has a surface area of 7.9209 Å$^2$ per HYLION-14. If any sphere has this surface area, the volume of it is 2.096 Å$^3$. This is only 0.001% of unit cell volume. (See supplementary information) Alkyl chains floating in $\Phi_{o,14}$ are difficult to act as a wall for gas adsorption, and only a very small amount of gas is absorbed. Therefore, huge space in $\Phi_{o,14}$ is ready for absorption. It has an inner space of 78 nm$^3$ through a pore with a 2 nm radius. This shows that strong absorption occurs with very strong nano-capillary force. Materials for water treatment have been designed with physical properties such as the polarity and pore size.[13] Therefore, it acted as sieves, and only surface properties were importantly treated. While these materials are passively operated with external energy for separation, nano-capilary force of $\Phi_{o,14}$ serves a new driving force to uptake solute itself.[8]

To confirm this, the contact angle (CA) of the various phases of HYLION-14 was measured. The water CA of soda lime glass is 65.1° **(Fig. S4)** and water CA of $\Phi_{m,14}$ is 106.7°. **(Fig. 4a)**. This is similar to water CA observed for the conventional long alkyl chain hydrocarbons such as wax[37]. However, water CA of $\Phi_{o,14}$ is 151.3°. **(Fig. 4b)** It demonstrates superhydrophobic surface $\Phi_{o,14}$ and it can be though as superhydrophobic surface based on their structural characteristics. Moreover, $\Phi_{o,14}$ exhibited a very strong adhesive property against water. Because the nano-capillary attraction occurs by the nanopore of $\Phi_{o,14}$. As a result, the water droplet was concretely pinned on the surface without any movement once the sample was upside down. **(Fig. 4c)** It verified that the surface of $\Phi_{o,14}$ may have liquid-solid area contact that makes high adhesive property.[6] On the other hand, ethanol CA of $\Phi_{m,14}$ is 23.1°.**(Fig. 4d)** This value conforms to usual ethanol CA of hydrophobic surfaces[38]. Notably, as a characteristic feature of the $\Phi_{o,14}$, no ethanol droplet is formed on the surface. Instead, rapid wetting occurs as can be seen in. **(Supplementary video 1)**

Thermogravimetric analysis (TGA) demonstrates the extremely high absorption capacity of the $\Phi_{o,14}$. **(Fig. 5a)** The absorbed amount was 92.6 wt.% for ethanol and 98 wt.% for 1,4-dioxane. $\Phi_{o,14}$ maintained their structure after soaking up with the guest molecules, as confirmed by PXRD patterns. **(Fig. 1c)** With the dropwise addition of the solvent to $\Phi_{o,14}$, the PXRD peak at 19.48° was gradually broadened, indicating twisting of HYLION-14 along the z-axis, while the other peaks remained unchanged during the soaking process[39].

Moreover, the rate of absorption of various organics by $\Phi_{o,14}$ is measured by modified edge immersion tests[40]. 1,4-dioxane, acetone, butanol, and ethanol are absorbed in 1.5 seconds **(Fig. 5b)** and the maximum velocity ($V_{max}$) of absorption was calculated using the hill function.[41] The $V_{max}$ values for 1,4-dioxane, acetone, butanol, and ethanol were 95.32, 98.68, 96.19, and 93.81, respectively. **(Fig. S5 – S8)**

Gas chromatography was carried out to verify high separation capability of $\Phi_{o,14}$ for an aqueous azeotropic mixture. It should remove 1,4-dioxane, acetone, butanol and ethanol from the water mixtures. The results of gas chromatograph markedly show the peaks of water and individual compounds in the azeotropic mixtures. **(Fig. 5c)** The profile of each sample showed a water peak at 2.5s, and the signals in order of acetone, butanol, ethanol and 1,4-dioxane were detected thereafter, respectively. $\Phi_{o,14}$ was well packed into a porous metal pipe and immersed in an agitated azeotropic mixture for 1 minutes. There were no peaks of organic compounds in the results of gas chromatography from the treated azeotropic mixtures. **(Figure 5d)** This demonstrates the huge potential of $\Phi_{o,14}$. Above all, the mixture of water and 1,4-dioxane is one of the most notorious positive azeotropes because of its infinite miscibility and similar boiling point. Nevertheless, the molecular metamaterial perfectly separates them without any kind of energy consumption. Moreover, acetone, butanol and ethanol azeotropes are also very important mixtures that must be separated in the biofuel industry. The selective process for separating acetone, butanol and ethanol from water is still the major bottleneck for the biofuel industry and the most expensive procedure regarding energy efficiency and productivity[42]. Herein, $\Phi_{o,14}$ can remove ABE from water spontaneously. Meta-separation can provide a cheap, convenient and complete separation method without energy consumption.

**Conclusions**

A novel strategy of separation exploiting the structural properties, rather than polarity, of a metamaterial was demonstrated. The regular and periodic structure of the molecular metamaterial provides low surface energy and super-hydrophobicity with high adhesive forces against water, whereas the nanopores confer a capillary effect for absorption of organics. The molecular metamaterial can uptake various organic solvents from 92 wt.% to 98 wt.%. Based on this exotic concept, azeotropes and various organic-water mixtures that are difficult to separate by pervaporation were easily separated in a few seconds. In particular, the separation using a capsule containing the molecular metamaterial opens up new possibilities for water treatment or organic solvent capture anywhere without large facilities. This approach has the potential to lead to great innovations, not only in the biofuel industry, but also in the elimination of various endocrine disruptors from environmental system.

Figures

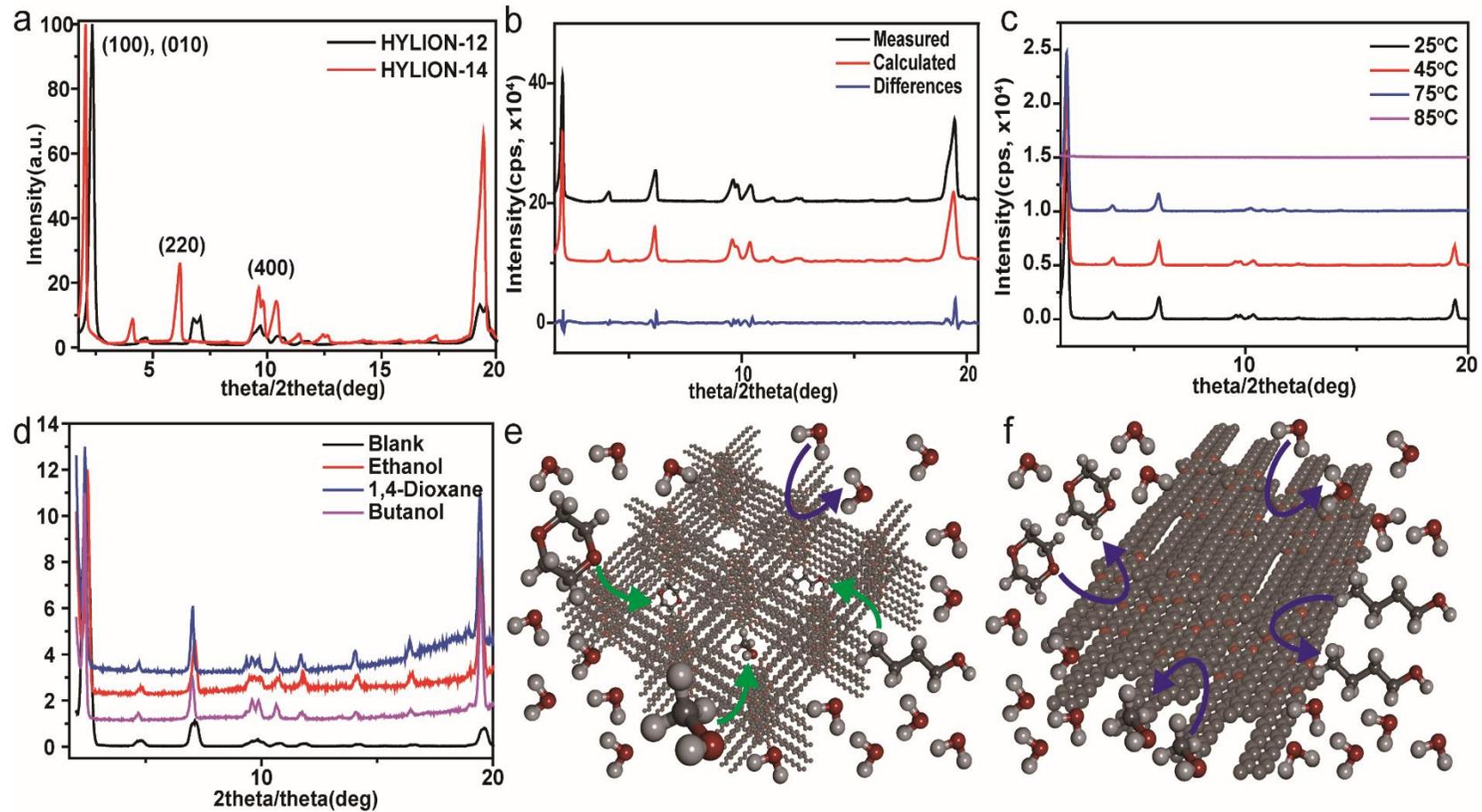

**Figure 1. Structural analysis of crystals of HYLION-14. a.** PXRD patterns of orthorhombic phase of HYLION-12 and HYLION-14 which are Dirac metamaterials **b.** Rietveld analysis of orthorhombic phase of HYLION-14. **c.** High-temperature PXRD patterns of $\Phi_{o,14}$ from 25°C to 85°C. **d.** PXRD patterns of $\Phi_{o,14}$ soaked with each solvent showing that all structural frameworks were maintained without disassembly **e.** Structure modeling of $\Phi_{o,14}$. It clearly shows porous structure of it **f.** Structure modeling of $\Phi_{m,14}$

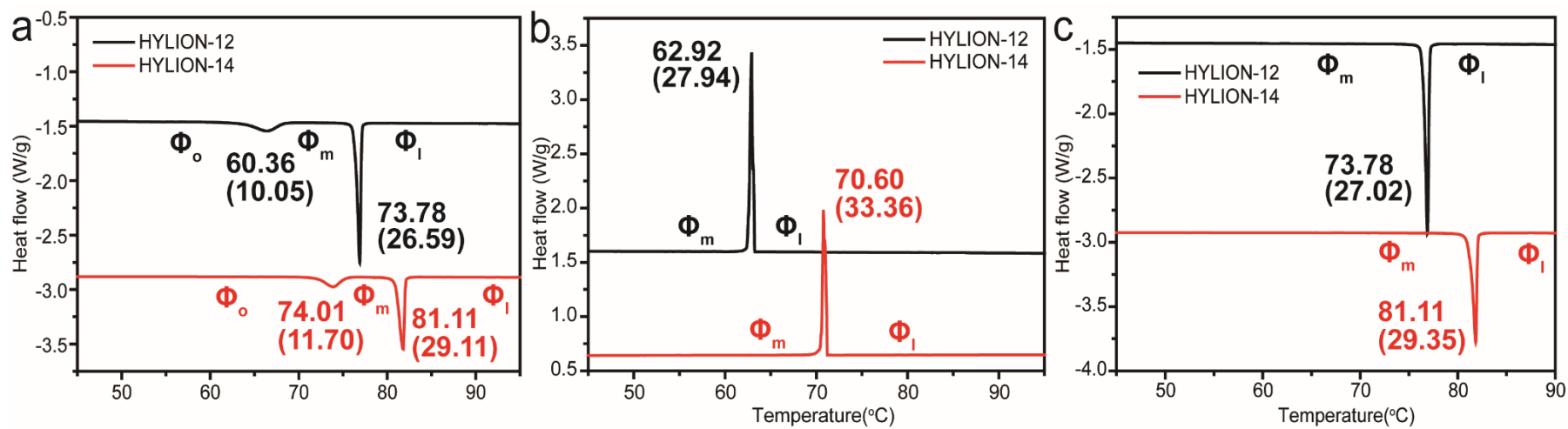

**Figure 2. DSC curves of the Φ₀ phases of HYLIONs-12 and -14.** Phase transition temperatures(°C) and associated enthalpy changes in brackets (kcal mol⁻¹) are represented. **a.** 1st heating cycle, **b.** 1st cooling cycle, **c.** 2nd heating cycle.

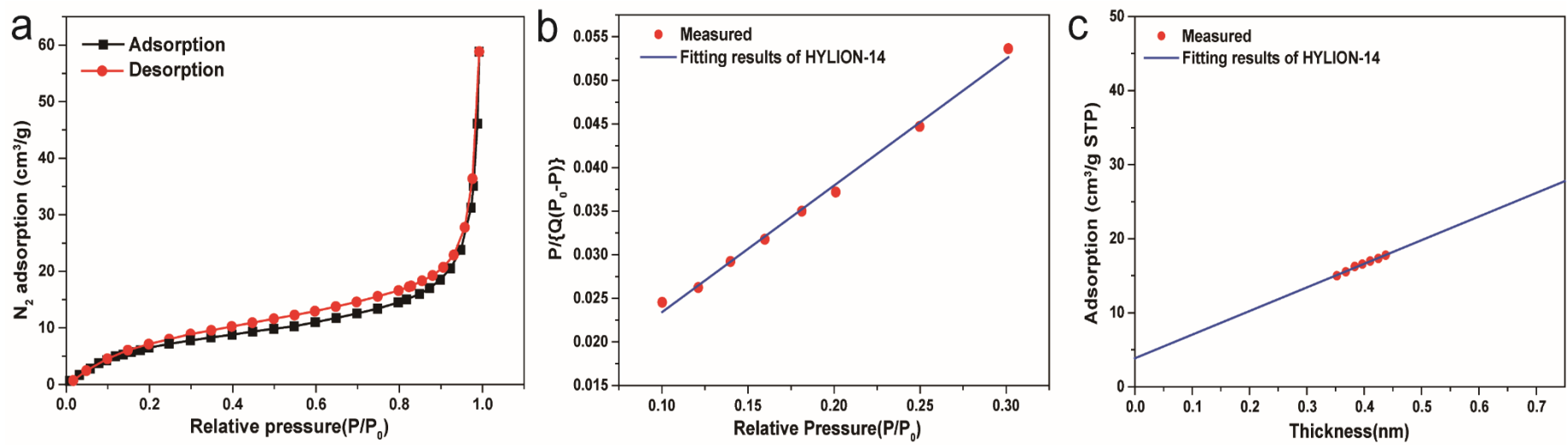

**Figure 3. The BET analysis of surface area and porosity of Φ$_{o,14}$ phases a.** N$_2$ adsorption-desorption isotherms of Φ$_{o,14}$. **b**. BET surface area. **c**. *t*-plot analysis.

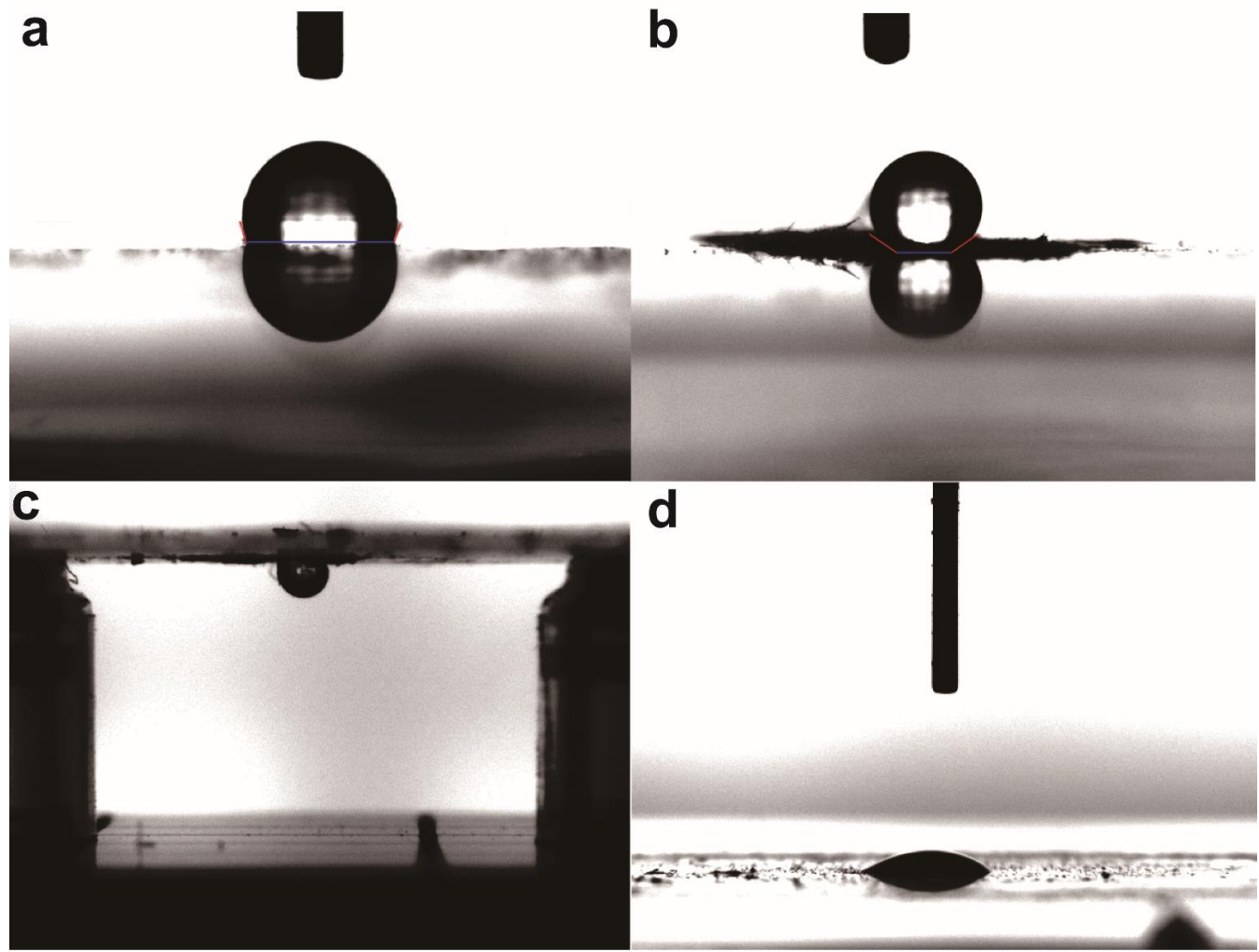

**Figure 4. Contact angles and wettability of the different HYLION-14 phases. a and b.** A drop of water on $\Phi_{m,14}$ (a) and $\Phi_{o,14}$(b). **c.** Inversed sample of drop of water $\Phi_{o,14}$ **d.** A drop of ethanol on $\Phi_{m,14}$.

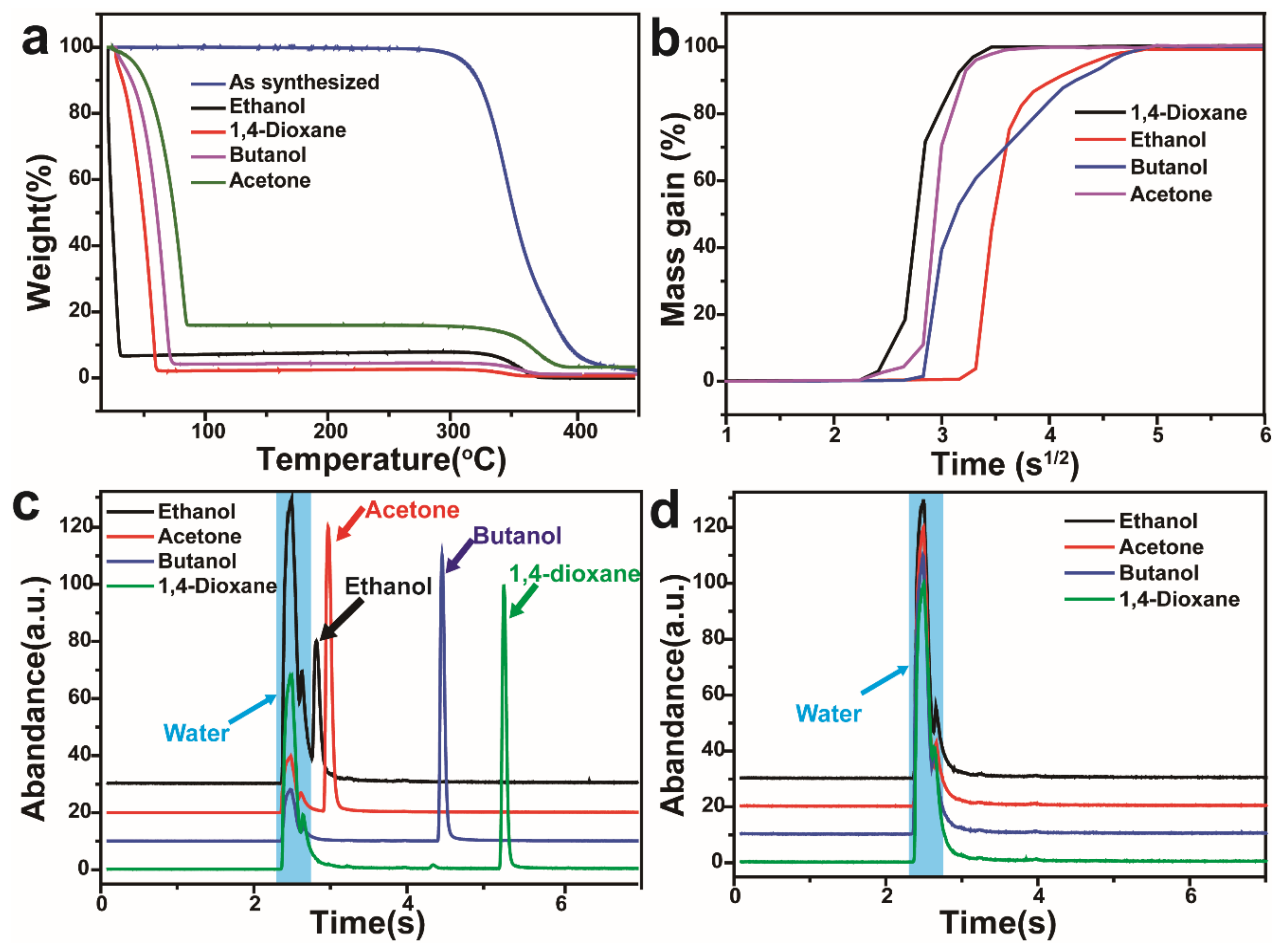

**Figure 5. Absorption and separation capability of $\Phi_{o,14}$ a.** TGA curves of pristine (blue) and $\Phi_{o,14}$, 1,4-dioxane (red), acetone (green), butanol (purple), and ethanol (black). **b.** Mass gain over time of exposure for 1,4-dioxane (black), acetone (purple), butanol (blue), and ethanol (red). **c.** Gas-chromatography(GC) curves of azeotropic mixtures with the 100 μg L$^{-1}$ concentration of ethanol (black), acetone (red), butanol (blue), and 1,4-dioxane (green) in water **d.** GC curves measured immediately after immersing $\Phi_{o,14}$ in azeotrope mixtures.

**Supporting information**

**Meta-separation: complete separation of organic-water mixtures by structural property of metamaterial.**

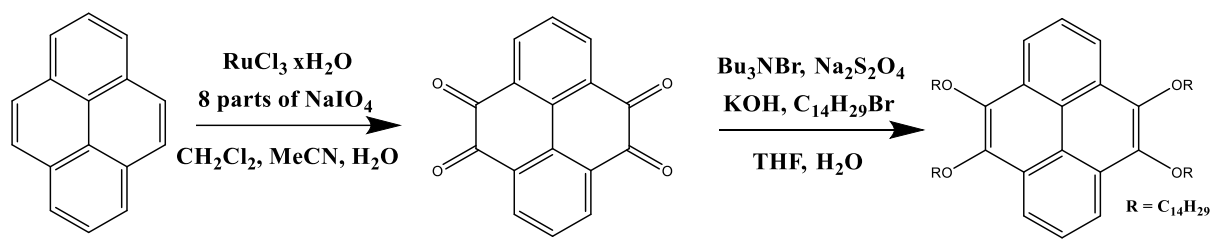

**Figure S1.** Synthesis of 4,5,9,10-tetrakis(tetradodecyloxy)-pyrene.

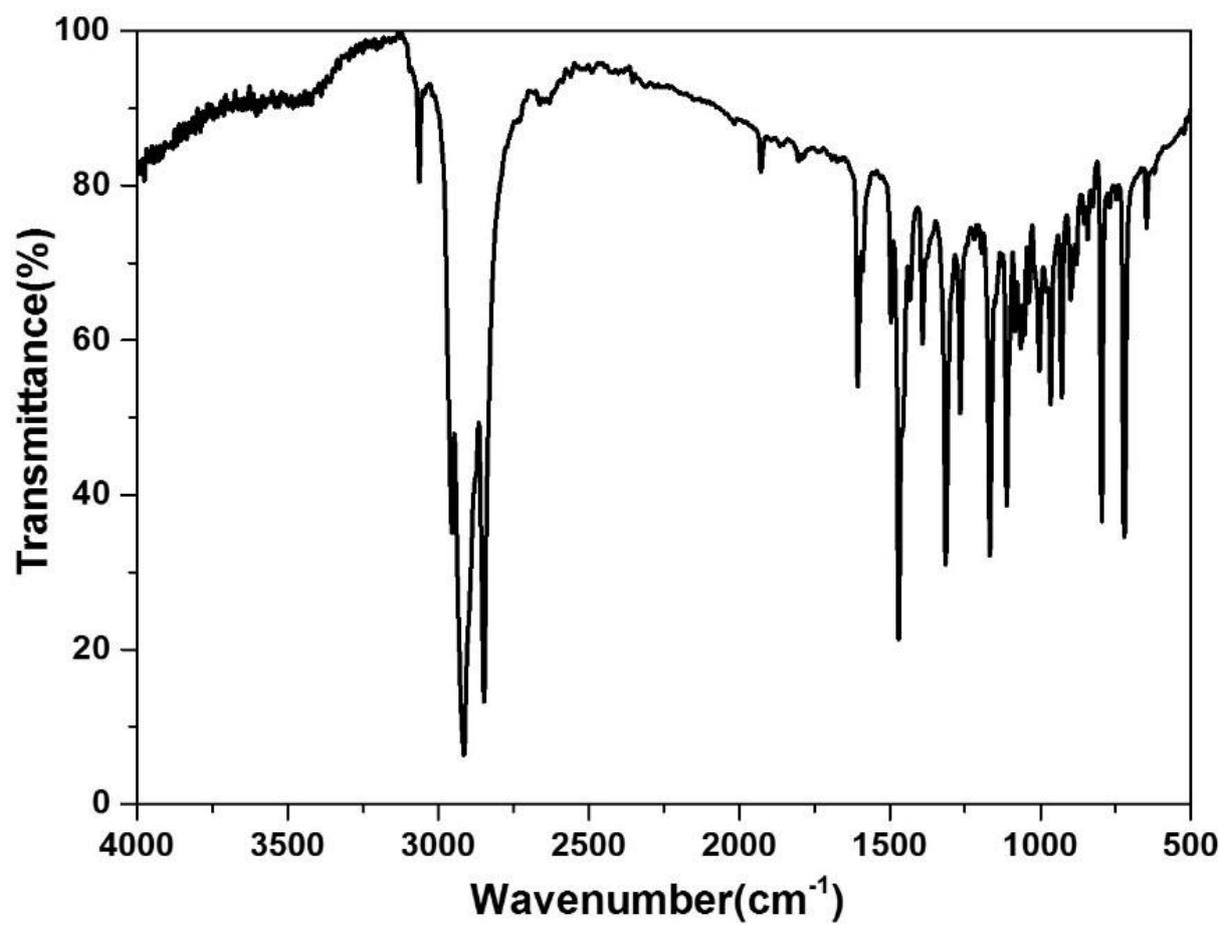

**Figure S2.** Infrared spectra of 4,5,9,10-tetrakis(tetradodecyloxy)-pyrene.

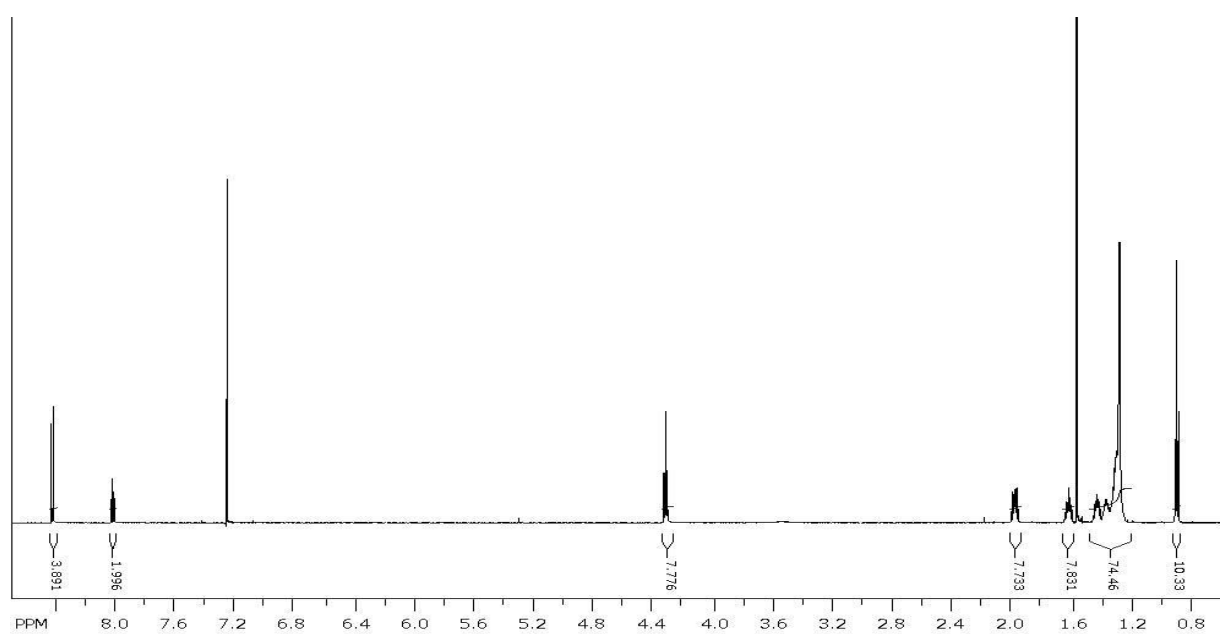

**Figure. S3** The NMR characterization of HYLION-14. 1H NMR (600MHz, CDCl3) δ 8.32(d, 4H), 7.71(t, 2H), 4,21(t, 8H), 1.91(m, 8H), 1.57(m, 8H), 1.40-1.27(m, 80H), 0.88(t, 12H).

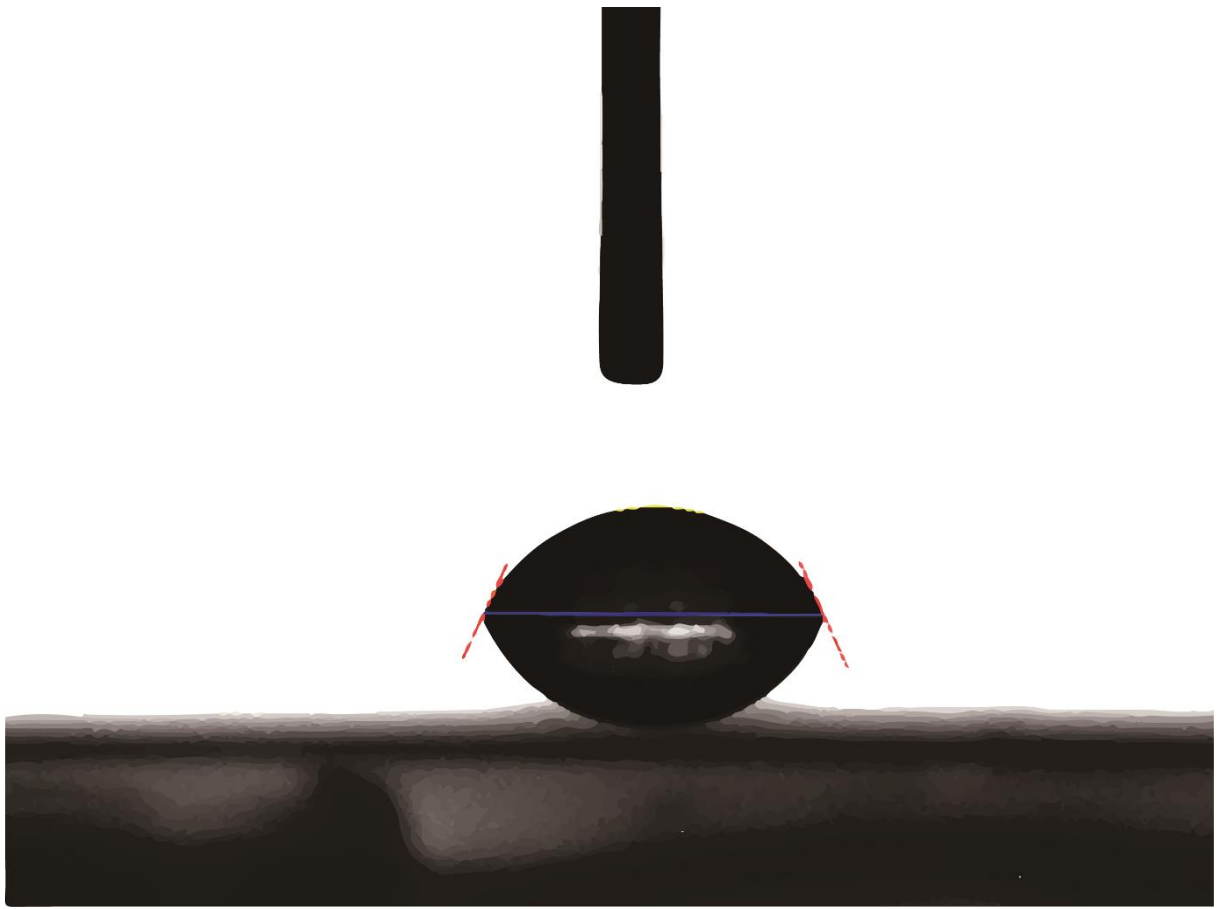

**Figure S4.** Contact angles a drop of water on bear soda-lime glass

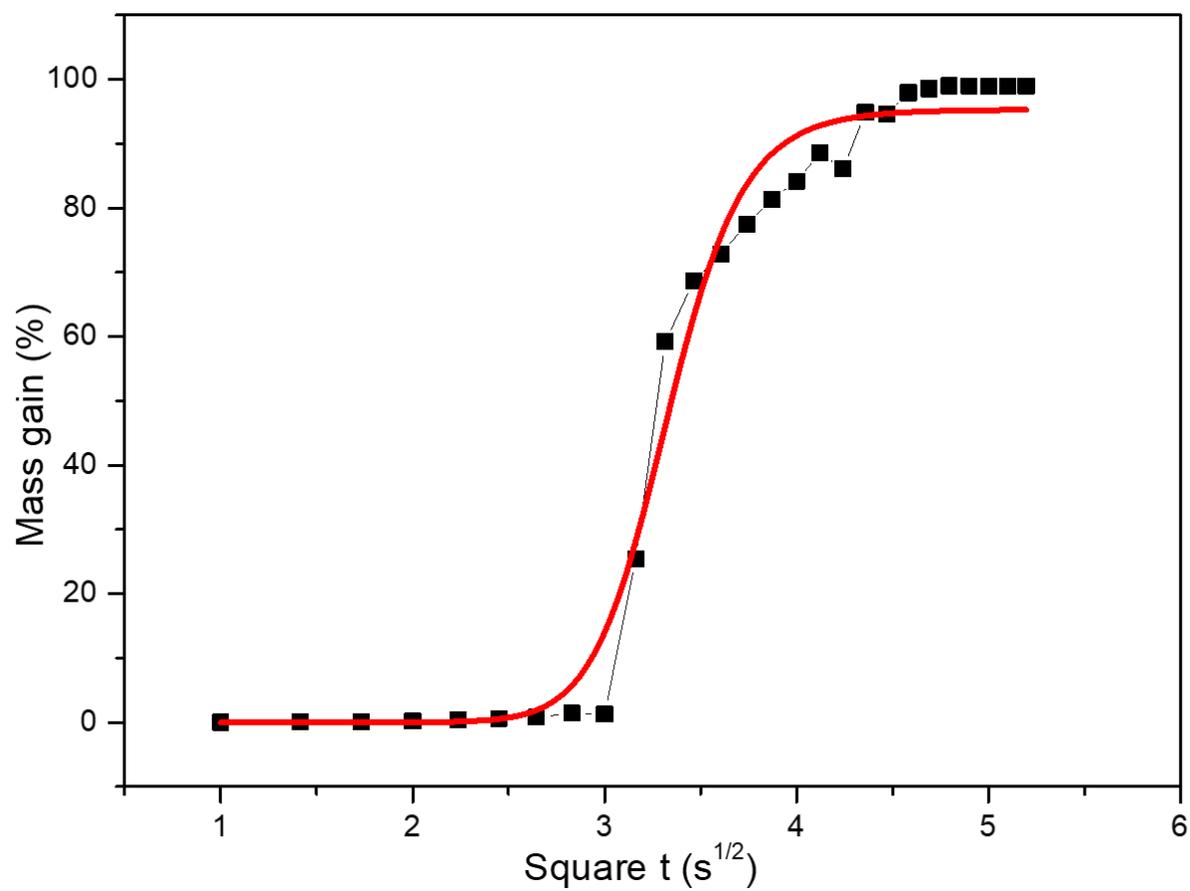

**Figure S5.** The edge immersion test graph which is fitted by Hill function for 1,4-dioxane.

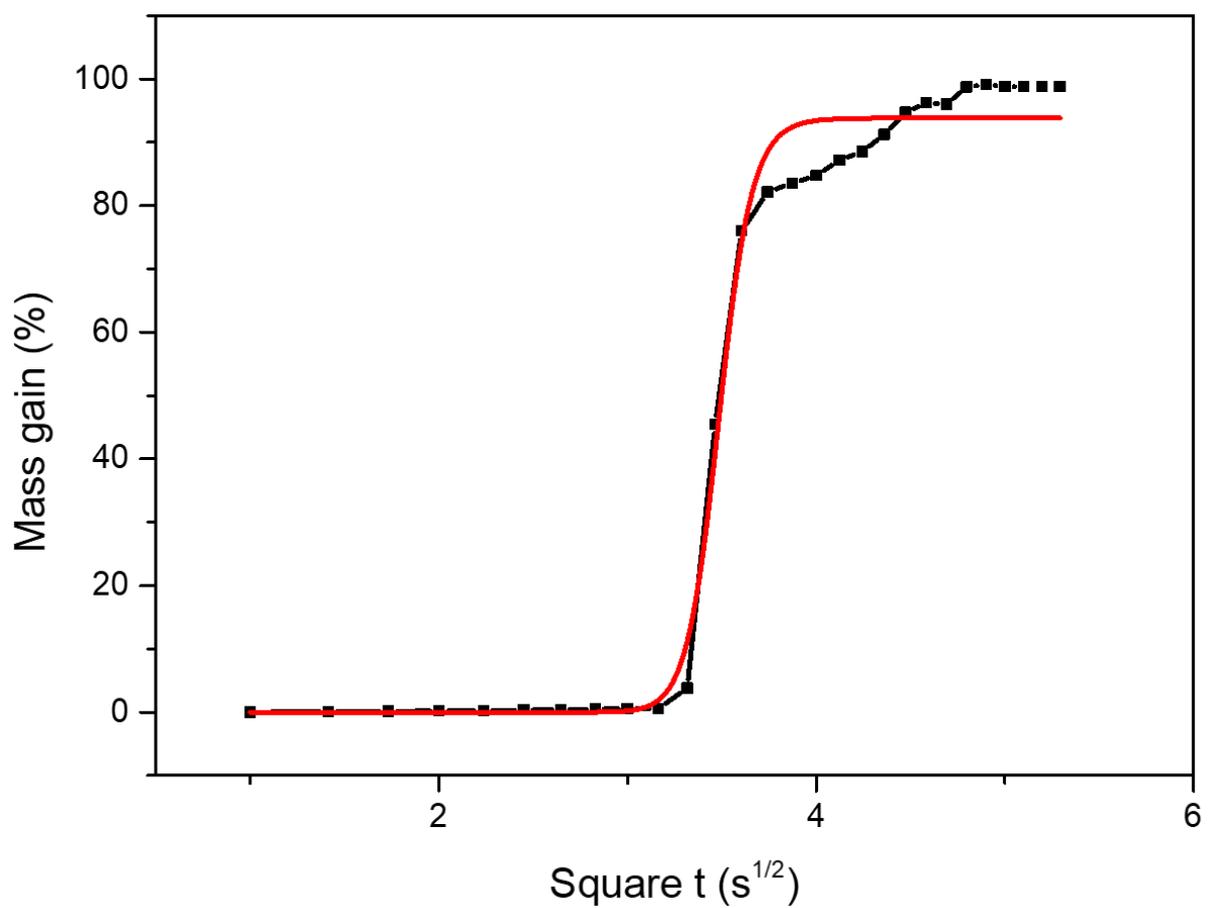

**Figure S6.** The edge immersion test graph which is fitted by Hill function for ethanol.

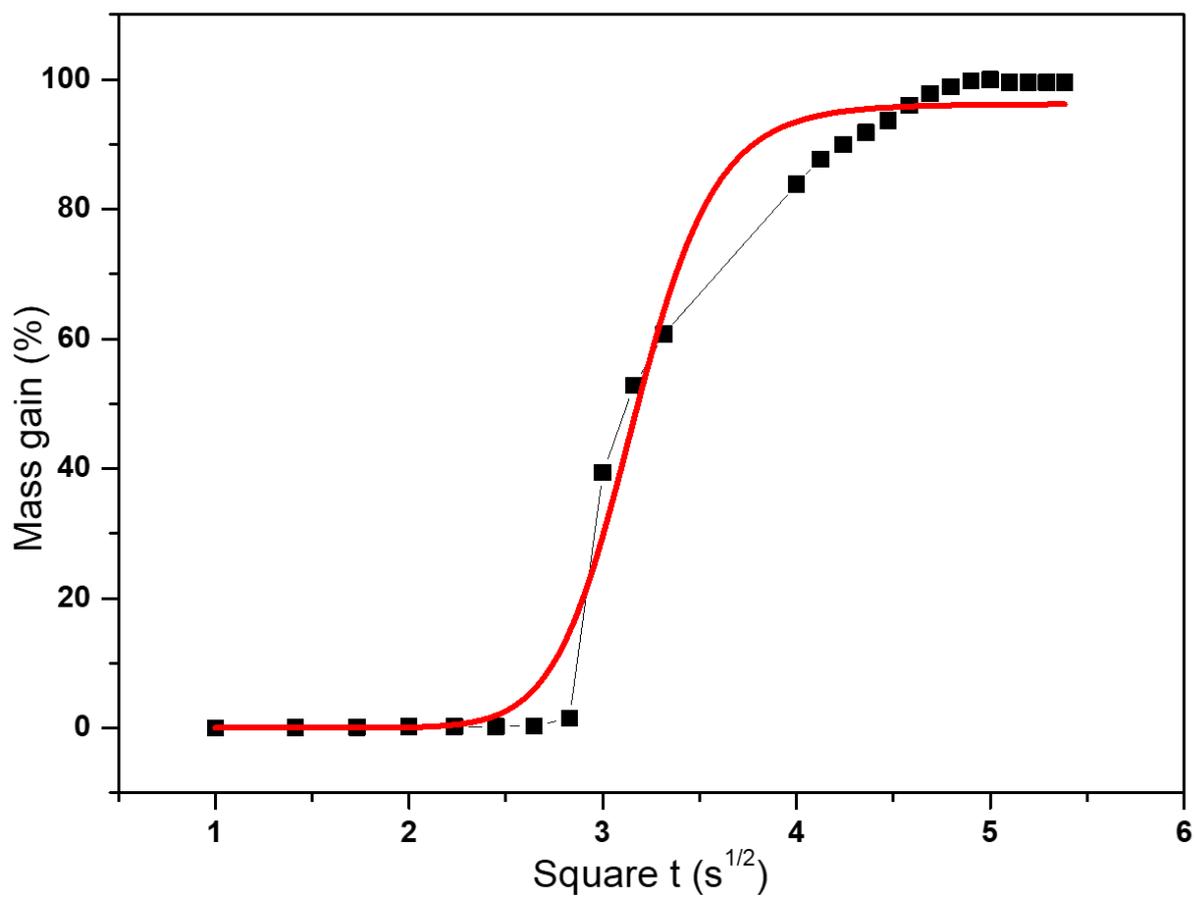

**Figure S7.** The edge immersion test graph which is fitted by Hill function for Butanol.

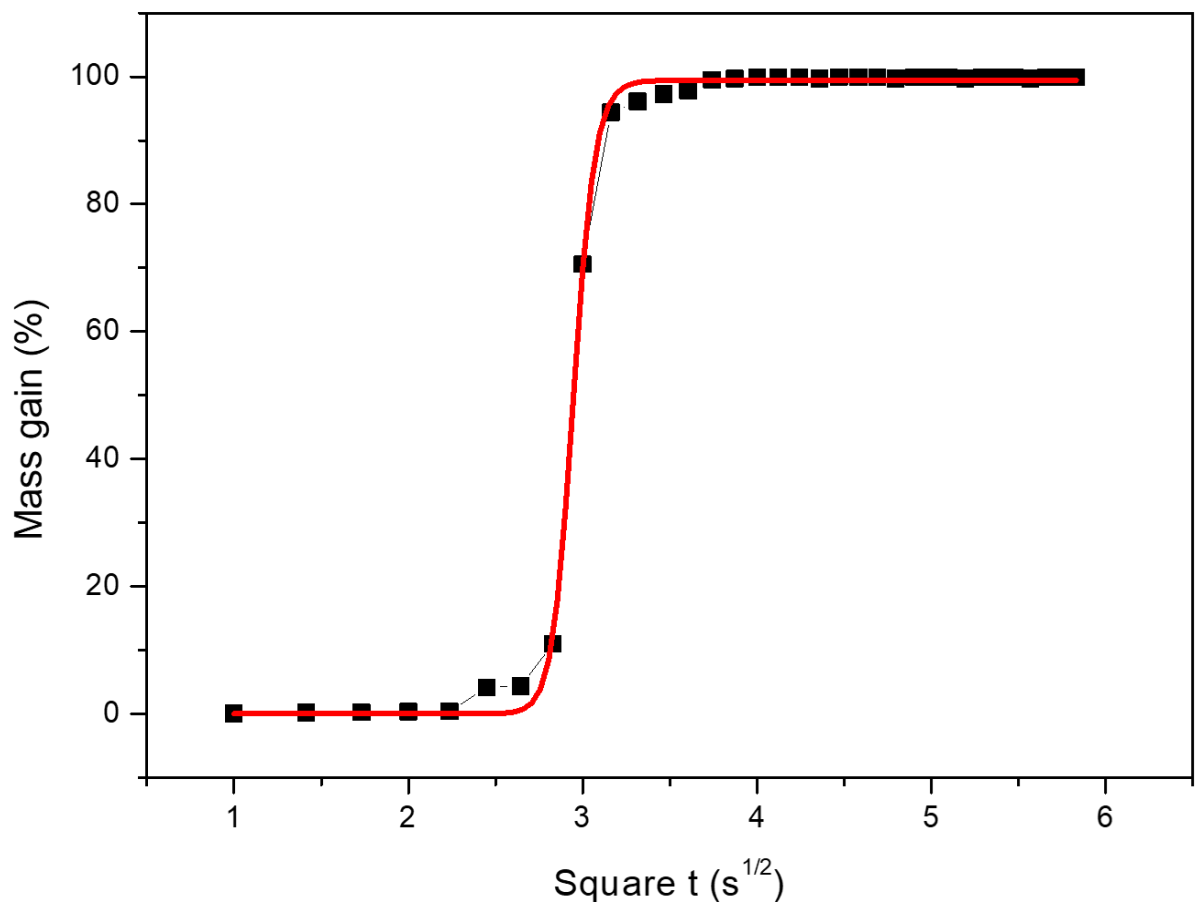

**Figure S8.** The edge immersion test graph which is fitted by Hill function for acetone.